\documentclass[runningheads]{llncs}
\usepackage[T1]{fontenc}
\usepackage{graphicx}
\usepackage{amsmath}
\usepackage{amssymb}
\usepackage[colorlinks=true,linkcolor=blue]{hyperref}
\usepackage{color}

\begin{document}
\title{Determination of galaxy photometric redshifts using Conditional Generative Adversarial Networks (CGANs)}
\titlerunning{Determination of photometric redshifts using CGAN.}
\author{M. Garcia-Fernandez\orcidID{0000-0001-8262-7541}}
\authorrunning{M. Garcia-Fernandez}
\institute{School of Architecture, Engineering and Design. Universidad Europea de Madrid. Calle Tajo s/n, Villaviciosa de Odon, 28670, Madrid, Spain
\email{manuel.garcia2@universidadeuropea.es}}
\maketitle              % typeset the header of the contribution
\begin{abstract}
Accurate and reliable galaxy redshift determination is a key requirement for wide-field photometric surveys. The estimation of photometric redshifts for galaxies has traditionally been addressed using artificial intelligence techniques trained on calibration samples, where both photometric and spectroscopic data are available. In this paper, we present the first algorithmic approach for photometric redshift estimation using Conditional Generative Adversarial Networks (CGANs). The proposed implementation is capable of producing both point estimates and probability density functions for photometric redshifts. The methodology is tested on Year 1 data from the Dark Energy Survey (DES-Y1) and compared against the current state-of-the-art Mixture Density Network (MDN) algorithm. The CGAN approach achieves comparable quality metrics to the MDN, demonstrating its potential and opening the door to the use of adversarial networks in photometric redshift estimation.

\end{abstract}
\section{Introduction}
Wide-field photometric surveys have been a major source of experimental results in Observational Cosmology. Among the many present and future photometric surveys, we find: DES\footnote{https://www.darkenergysurvey.org}, LSST\footnote{https://www.lsst.org}, PAU\footnote{https://pausurvey.org}, J-PAS\footnote{https://www.j-pas.org}, and {\it Euclid}\footnote{https://www.euclid-ec.org}. One of the key aspects of wide-field photometric surveys is the reliable determination of galaxy redshifts. In photometric surveys, the redshift of galaxy spectra is inferred by measuring the brightness of galaxies in broad-band filters, rather than determining the Doppler shift of their spectra using a high-resolution spectrometer.

The usual approach for translating brightness measured in broad-band filters into redshift is the use of artificial intelligence. These techniques utilize a calibration sample of galaxies with both known photometry and high-resolution spectra. This calibration sample is used by artificial intelligence algorithms to identify and discover patterns that relate brightness in different broad-band filters to spectroscopic redshift.

Previous artificial intelligence algorithms capable of determining point estimates for photometric redshift have included: neural networks \cite{2023arXiv231016304C,HOYLE201634,2024arXiv240909981M}, boosted decision trees \cite{2010ApJ...715..823G}, convolutional neural networks \cite{2018A&A...609A.111D,2021A&A...651A..55S}, Bayesian neural networks \cite{LIMA2022100510}, random forests \cite{10.1093/mnras/stad3976}, recurrent neural networks \cite{2024MNRAS.535.1844L}, and nearest neighbours \cite{2018AJ....155....1G}. Other algorithms capable of providing a probability density estimation for redshift include: classification algorithms \cite{10.1093/mnras/stv1567}, hierarchical models \cite{Leistedt_2019}, and mixture density networks \cite{zansari}. A comprehensive systematic review of the different types of photometric redshift algorithms can be found in \cite{2022ARA&A..60..363N,2019NatAs...3..212S}.

Current state-of-the-art photometric redshift estimation methods include Convolutional Neural Networks (CNNs) --which rely on both estimated galaxy magnitudes and raw imaging-- \cite{wei2025photometricredshiftestimationemission}, Bayesian Neural Networks (BNNs) --which rely on an explicit shape for the posterior distribution, usually Gaussian-- \cite{jones2024improvingphotometricredshiftestimation}, and Mixture Density Networks (MDNs) --which rely on a predefined mixture of probability densities-- \cite{zansari}.

A new class of neural networks that has not yet been explored for photometric redshift estimation is Generative Adversarial Networks (GANs) \cite{10.5555/2969033.2969125}. A particularly relevant variant of GANs that could potentially be used for photometric redshift determination is Conditional Generative Adversarial Networks (CGANs), which, instead of modeling the full probability density function of the underlying data, model the probability density function conditioned on some input \cite{2014arXiv1411.1784M}. A major advantage of CGANs over BNNs and MDNs is that they do not require any assumptions about the functional shape of the probability density function of the inferred photometric redshift.

In this paper, we propose the use of CGANs for estimating photometric redshifts using magnitudes measured in broad-band filters. This algorithm is tested with data from the Dark Energy Survey Y1, overlapping with SDSS Stripe-82 spectroscopic data. Results obtained by the proposed CGAN are compared with the current state-of-the-art MDN approach \cite{zansari}, which produces more general probability densities than BNNs. A comparison with CNN-based approaches is excluded, as the nature of the input data required for CNNs (imaging) differs from that required by CGANs, MDNs, or BNNs (photometric magnitudes).

\section{Methodology}
\subsection{Conditional Generative Adversarial Network}

Let ${\bf x}_i$ be a sample of photometric data from magnitudes measured in broad-band pass filters for the $i$-th galaxy in a dataset, and let $y_i$ be its corresponding known spectroscopic redshift, such that $\{({\bf x}_i, y_i)\}_{i=1}^{N_{train}}$ constitutes the training dataset. Let ${\bf z}_i$ be a vector of randomly generated numbers\footnote{Note that in the Computer Science literature on GANs, random vectors are typically denoted by $z$. This should not be confused with the redshift of galaxies, which in this paper is denoted as $y$ or $\hat y$ for spectroscopic and photometric redshifts, respectively. In contrast, Astronomy literature typically denotes redshift as $z$.}, associated with the $i$-th galaxy. The generator network $G$ is a function such that $\hat y_i = G({\bf z}_i|{\bf x}_i; \theta_G)$, where $\hat y_i$ is the estimated photometric redshift for the set of magnitudes ${\bf x}_i$, and $\theta_G$ are the weights defining the generator neural network. On the other hand, the discriminator network $D$ provides a function such that $p_i = D(\hat y_i; \theta_D)$, with $p_i \in [0, 1]$, acting as a classifier that determines whether $\hat y_i$ is real data from the training sample or synthetic data produced by the generator. Here, $\theta_D$ are the weights defining the discriminator network.

Training a GAN network constitutes a min-max optimization problem such that $\min_{\theta_G}\max_{\theta_D} V(D,G)$. The choice of the function $V(D,G)$ is a broad topic in Computer Science. As demonstrated by \cite{NIPS2016_cedebb6e}, any GAN can be interpreted as a special case of variational divergence estimation. Thus, the function $V(G,D)$ can be expressed in the most general form as
\begin{equation}
\begin{split}
V(D,G) = \mathbb{E}_{{\bf x}\sim p_d({\bf x})}[\mathbb{E}_{y\sim p_d(y)}[g_f(D(y|{\bf x}))] +\\ \mathbb{E}_{{\bf z}\sim p_z({\bf z})}[-f^*(g_f(D(G({\bf z}|{\bf x})|{\bf x})))]],
\end{split}
\end{equation} where $g_f$ denotes the output activation function and $f^*$ is the corresponding Fenchel conjugate of $g_f$ \cite{hiriart}. The functions $g_f$ and $f^*$ can be freely chosen, provided they are derived from an $f$-divergence \cite{CIT-004,1705001,NIPS2007_72da7fd6,JMLR:v12:reid11a}. The expectations $\mathbb{E}_{{\bf x}\sim p_d({\bf x})}$, $\mathbb{E}_{y\sim p_d(y)}$, and $\mathbb{E}_{{\bf z}\sim p_z({\bf z})}$ denote expectations over ${\bf x}$, $y$, and ${\bf z}$, respectively.

Accordingly, the loss function to be minimized for the discriminator network is given by
\begin{equation}
\mathcal{L}_D(\theta_D) = \frac{-1}{N_{batch}}\sum\limits_{i=1}^{N_{batch}} \left[g_f(D(y_i|{\bf x}_i;\theta_D)) - f^*(g_f(D(G({\bf z}_i|{\bf x}i);\theta_G)|{\bf x}i ;\theta_D))\right],
\end{equation} while the loss function for the generator, to be minimized simultaneously, is given by
\begin{equation}
\mathcal{L}_G(\theta_G) = \frac{-1}{N_{batch}}\sum\limits_{i=1}^{N_{batch}} g_f(D(G({\bf z}_i|{\bf x}_i;\theta_G)|{\bf x}_i;\theta_D)).
\end{equation}

Among all possible $f$-divergences, we select the Kullback-Leibler divergence (KL-divergence), for which the corresponding $g_f(x)$ and $f^*(g_f(x))$ are given by \cite{NIPS2016_cedebb6e}: \begin{equation} g_f(x) = x,\quad f^*(g_f(x)) = e^{x-1}. \end{equation}

Using this $f$-divergence approach, the photometric redshift inferred by the generator network -given a fixed set of magnitudes ${\bf x}_i$- becomes a function of the random vector ${\bf z}_j$, such that \begin{equation} \hat y_i({\bf z}_j) = G({\bf z}_j|{\bf x}_i). \end{equation}

The proposed topology of the generator neural network consists of a sequence of three fully connected layers, where the first two layers are followed by a Batch Normalization layer and a ReLU activation function. The input layer has $4 + Z_{DIM}$ neurons and $G_{DIM}$ output neurons. The hidden layer consists of $G_{DIM}$ input and output neurons. The final layer has $G_{DIM}$ input neurons and a single output neuron. The input to the generator network is the vector ${\bf x}$ of four galaxy magnitudes and the random vector ${\bf z}$. The output is the inferred photometric redshift $\hat y$, which is a random number following the conditional probability distribution of the underlying spectroscopic redshift given a fixed set of magnitudes.

The discriminator neural network also consists of three fully connected layers, with the first two followed by a Batch Normalization layer and a ReLU activation function. The input layer has 5 neurons\footnote{This corresponds to 4 input magnitudes (${\bf x}i$) plus the redshift ($y_i$ or $\hat y_i$).} and $D_{DIM}$ output neurons. The hidden layer has $D_{DIM}$ input and output neurons. The final layer has $D_{DIM}$ input neurons and a single output neuron, followed by a sigmoid activation function, $\sigma(x) = 1/(1+e^{-x})$. The input to the discriminator is a redshift (either spectroscopic $y$ or photometric $\hat y$) and the vector ${\bf x}$ of galaxy magnitudes. The output is a decimal number between 0 and 1 indicating the probability that the input galaxy is real.

\begin{figure}
    \centering
    \includegraphics[width=\linewidth]{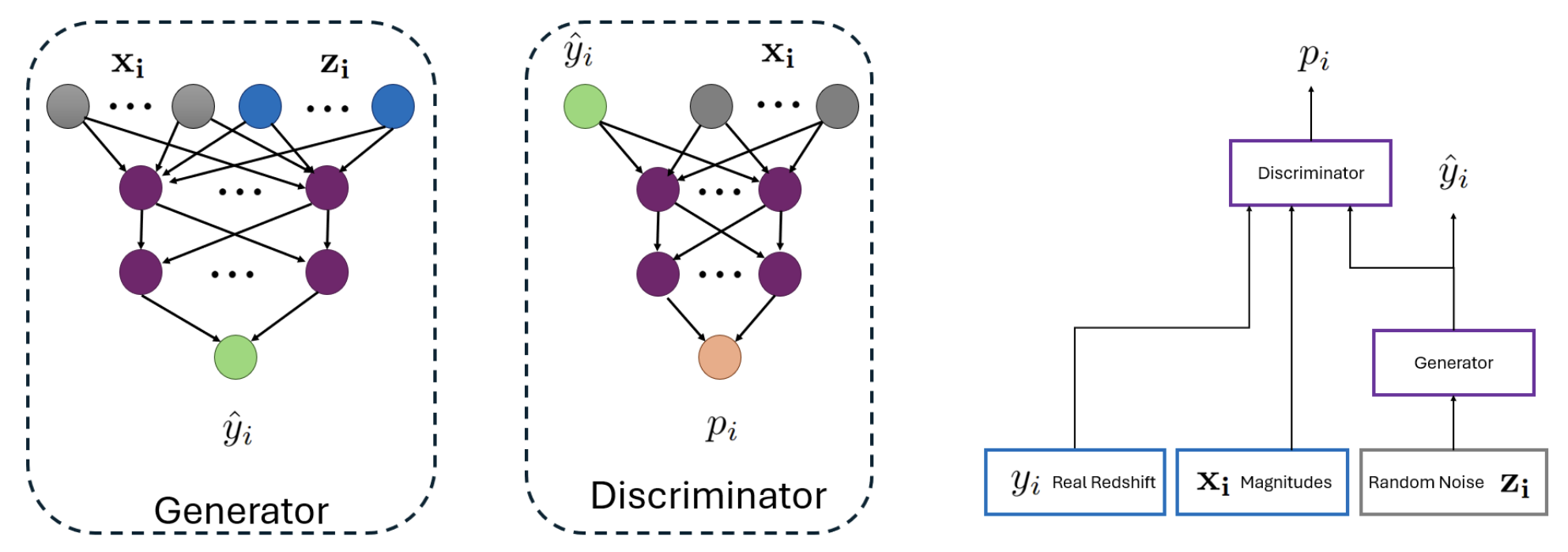}
    \caption{Left: Topology of the Generator and Discriminator networks in the CGAN model. Gray neurons ${\bf x}_i$ represent the input feature vector containing the four {\tt MAG\_AUTO} magnitudes. Blue neurons correspond to the input random vector ${\bf z}_i$. The green neuron $\hat y_i$ denotes the predicted photometric redshift. The orange neuron $p_i$ represents the probability that a given observation is synthetic data generated by the Generator or real data from the training sample. Purple neurons indicate the hidden layers. Right: Interaction between the Generator and Discriminator networks in the CGAN framework.}
    \label{fig:topology}
\end{figure}

More recent works on GANs have made extensive use of the Wasserstein formalism \cite{pmlr-v70-arjovsky17a} (WGANs) instead of $f$-divergences, demonstrating superior empirical performance and more stable convergence of GANs \cite{brock2018large,karras2018progressive}. Nevertheless, WGANs do not allow the interpretation of outputs as probability densities \cite{pmlr-v119-song20a}. Therefore, the Wasserstein formalism has been excluded from this work.

On the right side of \autoref{fig:topology}, the training process and the interaction between the Generator and Discriminator neural networks are illustrated. A tensor of random noise ${\bf z}_i$ (gray box) is fed into the Generator network, producing an estimate of the photometric redshift. Tensors containing real spectroscopic redshifts ($y_i$) and estimated photometric redshifts ($\hat y_i$) are fed into the Discriminator network, which attempts to determine which records correspond to real spectroscopic data and which are generated by the Generator network, given the magnitude data ${\bf x}_i$ (blue box).

\subsection{Mixture Density Network}
Mixture Density Networks (MDNs) are a type of machine learning model that combines a neural network with a parametric mixture model \cite{Bishop1994MixtureDN}. Instead of producing a point estimate, MDNs learn the conditional probability distribution as a linear combination of a finite set of individual probability distributions. To achieve this, the neural network in the MDN is trained to predict the parameters that characterize the component distributions, as well as the mixing coefficients. The most commonly used probability distribution is the Gaussian. Accordingly, the neural network aims to determine the set of parameters $\{(\mu_i, \sigma_i, \pi_i)\}_{i=0}^{N_g}$, where $\mu_i$ is the mean, $\sigma_i$ the standard deviation, and $\pi_i$ the mixing coefficient of the $i$-th Gaussian (with the constraint $\sum_{i=0}^{N_g} \pi_i = 1$). The number of Gaussians in the mixture, $N_g$, is a configuration parameter of the model that must be fixed in advance.

To enable comparison with a Mixture Density Network, in this work we adopt the MDN implementation proposed by \cite{zansari}, using the code provided in the companion GitHub repository\footnote{\url{https://github.com/ZoeAnsari/MixtureModelsForPhotometricRedshifts}}, which we ported from {\tt Keras} to {\tt PyTorch}. The configuration parameters of the MDN model are kept the same as in the original implementation, except for the number of neurons in the input layer, where we use 4 neurons instead of the original 22. This modification adapts the original code from \cite{zansari} to the 4 magnitude features in our dataset, as opposed to the 22 magnitude features used in the original work. The resulting MDN model consists of 30 Gaussian components, a fully connected input layer with 4 neurons, and a hidden fully connected layer with 22 neurons.

\section{Data Analysis}
The proposed CGAN was tested on DES-Y1 data matched with spectroscopic redshifts from SDSS galaxies. The data were obtained from the public DR1 release of the Dark Energy Survey Collaboration, available through the NCSA repository\footnote{\url{https://des.ncsa.illinois.edu/releases/y1a1}}.

The CGAN code was implemented in Python using {\tt PyTorch}. The full codebase for this analysis is available in the author's GitHub repository\footnote{\url{https://github.com/mgarciafernandez-uem/CGAN-photoz}}.

From the DES-Y1 dataset, we selected the Stripe-82 subset of galaxies with matched spectroscopic redshifts from SDSS. For sample selection, we restricted the dataset to galaxies with spectroscopic redshifts in the range $0.0 < z_{sp} < 0.8$. This quality cut was applied to exclude the long tail of galaxies extending up to redshift 2, which contains only a few objects. Including such underrepresented galaxies in the training sample could introduce biases, as the neural network might infer magnitude–redshift relationships from insufficient data. The selected photometric features are the {\tt MAG\_AUTO} magnitudes in the $griz$ band-pass filters. The final calibration sample consists of 33,410 galaxies. This sample was split into training, testing, and validation subsets, containing $80\%$, $10\%$, and $10\%$ of the galaxies, respectively. The training set is used to optimize the parameters of the photometric redshift algorithm, the test set is used to monitor overfitting during training, and the validation set is used to evaluate the model's performance on unseen data.

\begin{figure}
    \centering
    \includegraphics[width=0.75\linewidth]{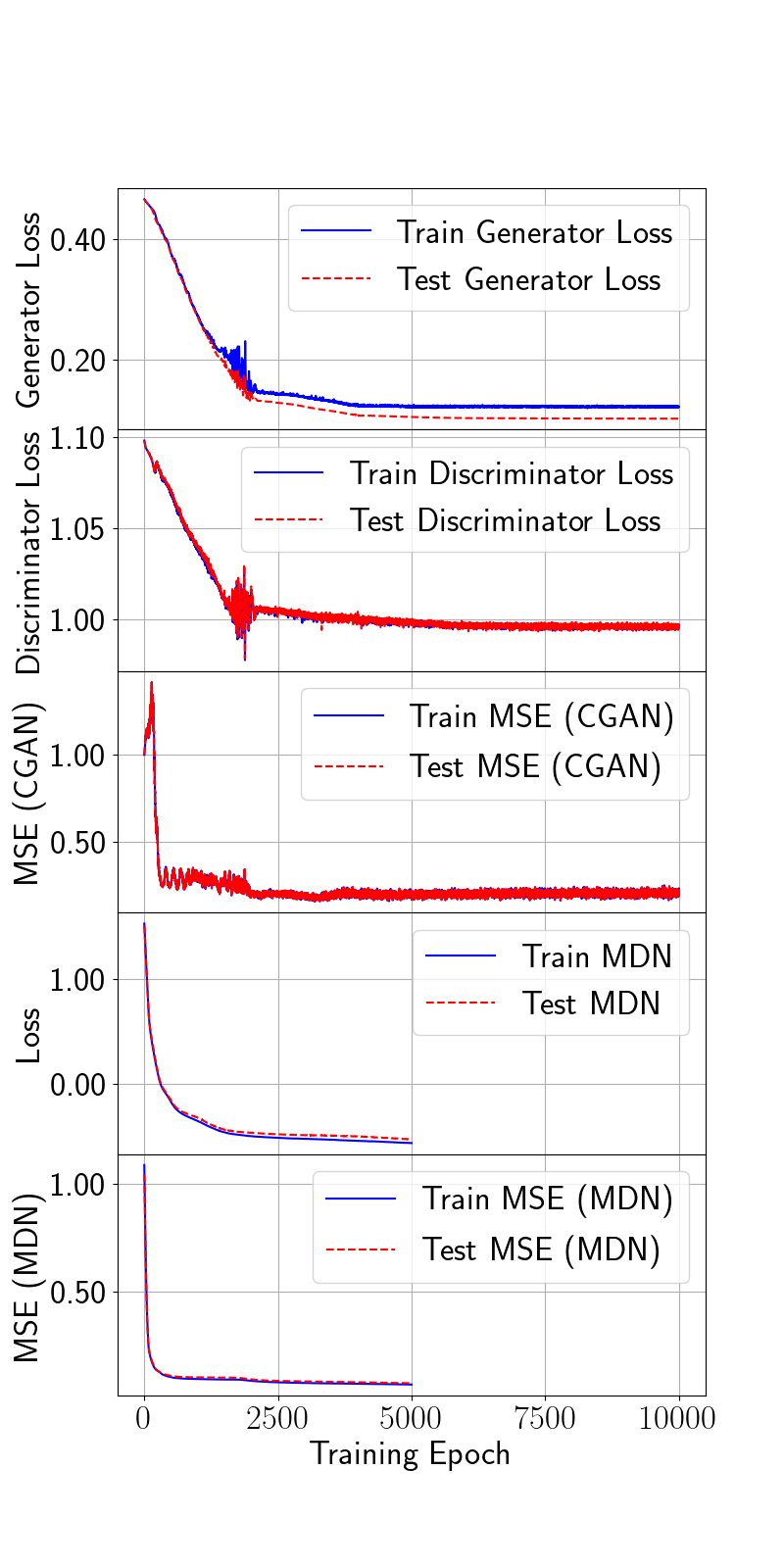}
    \caption{Loss functions and mean squared error (MSE) for CGAN and MDN. The top three plots show the loss functions and MSE for the CGAN, while the bottom two plots display the loss and MSE for the MDN. The loss functions for the CGAN's generator and discriminator networks have been shifted by +1 to avoid negative values on the Y-axis. The MDN training curves terminate earlier than those of the CGAN, as the MDN was trained for 5,000 epochs, compared to 10,000 epochs for the CGAN.}
    \label{fig:losses}
\end{figure}

The model hyperparameters defining the sizes of the dense layers and the random noise vector are $G_{DIM} = 32$, $D_{DIM} = 32$, and $Z_{DIM} = 20$. The training strategy employed the Adam optimizer with an initial learning rate of $lr = 10^{-4}$ for both the generator and discriminator networks. Training was conducted over 10,000 epochs, using a step learning rate schedule that reduced the learning rate by a factor of 0.2 every 2,000 epochs. The same training strategy was applied to both networks. The evolution of the generator and discriminator losses over the training epochs is shown in \autoref{fig:losses}, demonstrating stable and well-behaved convergence for both networks.

The MDN training process involves 5,000 training epochs with an initial learning rate of $lr = 10^{-4}$. The evolution of the loss and mean squared error (MSE) during training is shown in \autoref{fig:losses}.

\begin{figure}
    \centering
    \includegraphics[width=\linewidth]{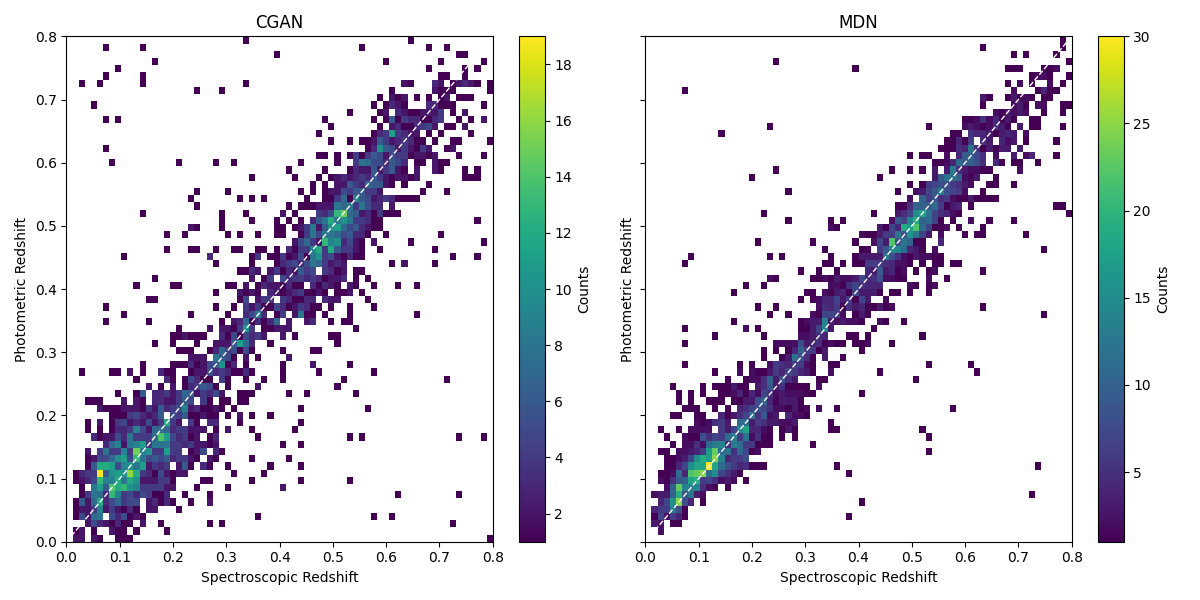}
    \caption{Comparison of photometric and spectroscopic redshifts of galaxies. Distributions are shown for the proposed Conditional Generative Adversarial Network (CGAN) approach and the Mixture Density Network (MDN). The white dashed line serves as a visual guide for the identity line.}
    \label{fig:pred_scatter}
\end{figure}

A comparison of the dispersion between the true spectroscopic redshift and the inferred photometric redshift is presented in \autoref{fig:pred_scatter}. From these plots, it is evident that the photometric redshifts inferred by both the proposed CGAN approach and the MDN accurately trace the true spectroscopic redshift distribution of the test sample of galaxies.

To evaluate the quality of both point estimates and probability density estimates, we follow the methodology proposed in \cite{TEIXEIRA2024100886}, where all metrics are computed over the validation sample of galaxies, which has not been seen by either the CGAN or the MDN during training.

\subsection{Point estimation quality metrics}

Point estimation quality metrics include the mean absolute bias ($\bar{|\Delta z|}$), the Normalized Median Absolute Deviation ($\sigma_{\mathrm{NMAD}}$), and the outlier ratio ($\eta$). These metrics are computed across 20 spectroscopic redshift bins within the interval $0 \le z_{sp} \le 0.8$.

The Normalized Median Absolute Deviation is defined as: \begin{equation} \sigma_{\mathrm{NMAD}} = 1.48 \times \mathrm{median}\left(\frac{\Delta z - \mathrm{median}(\Delta z)}{1 + z}\right), \end{equation} while the outlier ratio ($\eta$) is defined as the fraction of galaxies satisfying: \begin{equation} \left|\frac{\Delta z}{1 + z}\right| > 0.15. \end{equation}

Confidence intervals for these metrics are computed at the $95\%$ confidence level using bootstrapping. This is implemented by generating 1,000 bootstrap samples from the test dataset, each with the same number of galaxies as the original test set, sampled with replacement. For each bootstrap sample, the point estimation quality metrics are computed. The resulting distribution of 1,000 values for each metric is then used to determine the $2.5\%$ and $97.5\%$ quantiles, which define the lower and upper bounds of the confidence intervals.

\begin{figure}
    \centering
    \includegraphics[width=\linewidth]{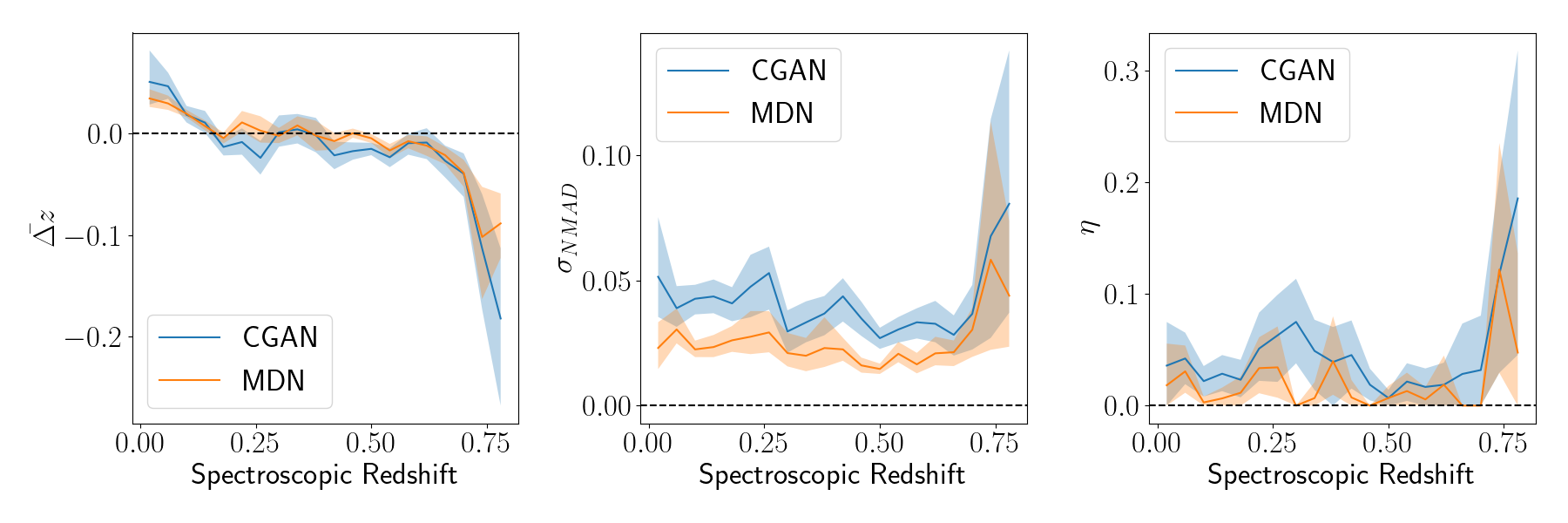}
    \caption{Point estimation quality metrics comparison for CGAN and MDN. The left panel shows the mean absolute bias $\bar{|\Delta z|}$, the center panel shows the Normalized Median Absolute Deviation $\sigma_{\mathrm{NMAD}}$, and the right panel shows the outlier ratio $\eta$. The solid lines represent the mean values per redshift bin, while the shaded areas indicate the $95\%$ confidence intervals, computed using bootstrapping.}
    \label{fig:point-quality}
\end{figure}

Visualizations of these quality metrics are shown in \autoref{fig:point-quality}. From these plots, we observe that both the CGAN and MDN models exhibit comparable performance across the metrics. However, the MDN demonstrates slightly higher accuracy, while the CGAN is more prone to outliers.

\subsection{Probability-density-function quality metrics}

Given a galaxy with spectroscopic redshift $z_{sp}$ and photometric redshift $z_{ph}$ with an associated probability density function (PDF) $\phi(z)$, its Probability Integral Transform (PIT) is defined as \cite{d5689415-e530-31d7-b8c7-8eb3cd1bfc8e,LIMA2022100510,polsterer2016uncertainphotometricredshifts}: \begin{equation} PIT = \int\limits_{-\infty}^{z_{sp}} \phi(z),dz. \end{equation} As stated by \cite{TEIXEIRA2024100886}, a properly calibrated PDF will produce a uniform distribution of PIT values over a large sample of galaxies.

The Odds metric \cite{LIMA2022100510} for a galaxy with PDF $\phi(z)$ and photometric redshift $z_{ph}$ is defined as: \begin{equation} Odds = \int\limits_{z_{ph}-\xi}^{z_{ph}+\xi} \phi(z),dz, \end{equation} where $\xi \in \mathbb{R}$ is a fixed parameter. Following \cite{TEIXEIRA2024100886}, we set $\xi = 0.06$. A distribution of Odds values skewed toward higher values indicates that the PDFs are narrow and concentrated around the most probable value, suggesting a reliable estimation. Conversely, low Odds values indicate broader PDFs.

The Coverage Test \cite{DALMASSO2020100362,hermans2020likelihoodfreemcmcamortizedapproximate} is computed by taking each galaxy’s spectroscopic redshift $z_{sp}$ and its PDF $\phi(z)$. For a given confidence level $1 - \alpha$, the symmetric interval $[z_l, z_u]$ enclosing a probability of $1 - \alpha$ is defined by: \begin{equation} \frac{\alpha}{2} = \int\limits_{-\infty}^{z_l} \phi(z),dz = \int\limits_{z_u}^{\infty} \phi(z),dz. \end{equation} The fraction of galaxies for which $z_l \le z_{sp} \le z_u$ should ideally equal $1 - \alpha$. A lower-than-expected fraction indicates that the PDFs are too narrow, suggesting overconfidence in the model. Conversely, a higher-than-expected fraction indicates overly broad PDFs, suggesting underconfidence.

\begin{figure}
    \centering
    \includegraphics[width=\linewidth]{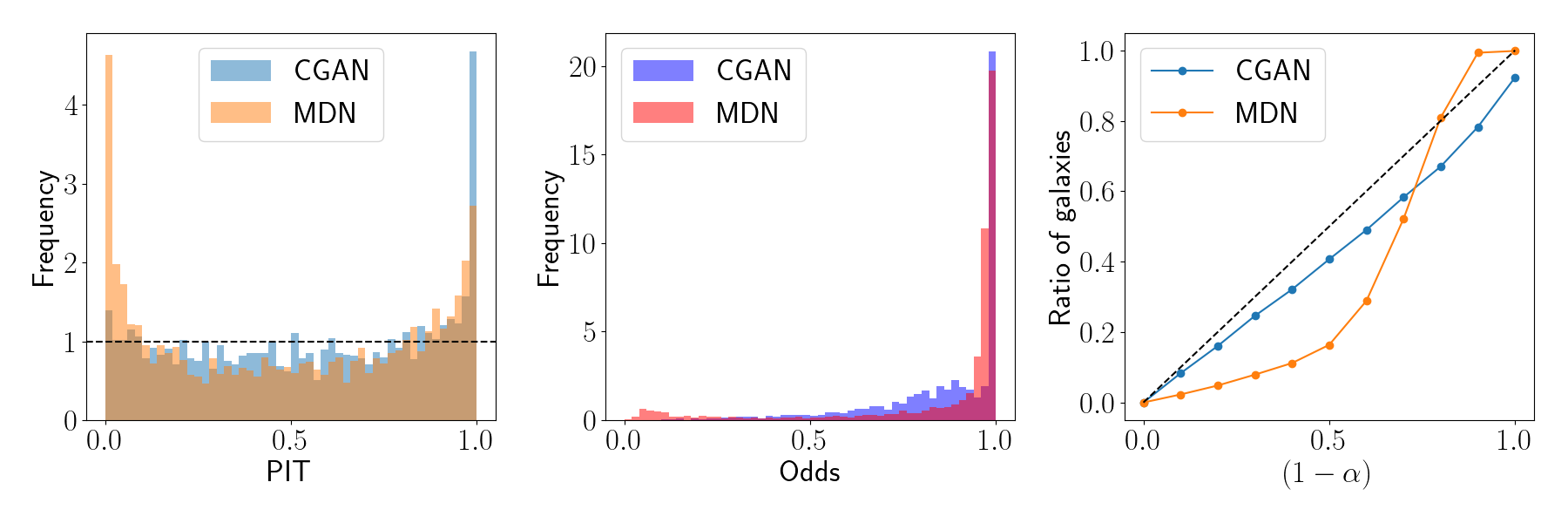}
    \caption{Probability density quality metrics comparison for CGAN and MDN. The left panel shows the Probability Integral Transform (PIT), the center panel shows the Odds distribution, and the right panel displays the credibility diagram. The black dashed lines represent the ideal case for perfectly calibrated probability density functions.}
    \label{fig:quality-pdf}
\end{figure}

Visualizations of these quality metrics are shown in \autoref{fig:quality-pdf}. From these plots, we observe that the PIT and Odds distributions for both the CGAN and MDN approaches are very similar. However, the Coverage Test reveals that the CGAN-generated PDFs are skewed toward overconfidence, as the measured coverage falls below the expected confidence levels.

\begin{figure}
    \centering
    \includegraphics[width=\linewidth]{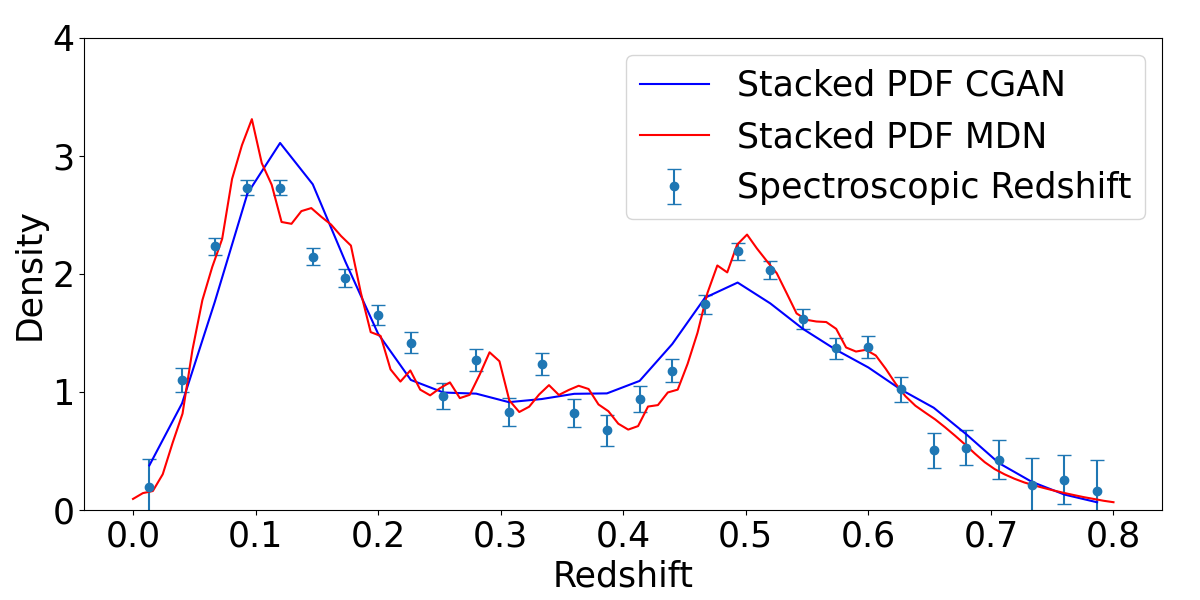}
    \caption{Comparison of the spectroscopic redshift distribution with the stacked probability densities inferred by CGAN and MDN. The plot compares the true distribution of spectroscopic redshifts with the distribution obtained by stacking the individual probability density functions predicted by the CGAN and MDN algorithms.}
    \label{fig:stacked-pdf}
\end{figure}

Additionally, to assess the ability of each algorithm to recover the underlying probability density function, we compare the distribution of the actual spectroscopic redshifts of the galaxies with the distribution obtained by stacking all individual probability density functions produced by each algorithm. The results are shown in \autoref{fig:stacked-pdf}, where we observe that the stacked probability densities provide a similar representation of the underlying redshift distribution. However, the MDN appears to more closely match the spectroscopic data than the CGAN.

\section{Conclusions}
In this paper, we presented a novel artificial intelligence technique for photometric redshift estimation using a Conditional Generative Adversarial Network (CGAN). The proposed CGAN approach was tested on Year 1 data from the Dark Energy Survey (DES-Y1) and compared against a Mixture Density Network (MDN).

Both point estimation and probability density quality metrics indicate that the CGAN performs comparably to the MDN. Although the MDN shows slightly better performance across most metrics, this work serves as a proof of concept, demonstrating that CGANs -when trained using the $f$-divergence formalism- are a viable alternative for photometric redshift estimation. This opens the door to further exploration of CGANs in this field.

A major advantage of the methodology proposed in this work, compared to state-of-the-art approaches such as MDNs and Bayesian Neural Networks (BNNs), is its ability to estimate the photometric redshift probability density without requiring an explicit parametric formulation.

The limitations of this study include the relatively small dataset and the use of a heterogeneous sample containing galaxies of various types, each known to exhibit different redshift probability distributions. Future work will explore more refined galaxy classifications (e.g., main-sequence galaxies, Luminous Red Galaxies) as an additional input parameter to the CGAN, computed in a preprocessing step. Further developments will also involve larger galaxy samples to enable more precise and robust measurements.

\begin{credits}
\subsubsection{\discintname}
No competing interests to declare.
\end{credits}
\bibliographystyle{splncs04}
\bibliography{bibliography}

\end{document}